\documentclass[cameraready]{Interspeech}

\title{audiobook-cc: Controllable Long-context Speech Generation for Multicast Audiobook}

\author[affiliation={1}]{Min}{Liu}
\author[affiliation={1}]{JingJing}{Yin}
\author[affiliation={1}]{Xiang}{Zhang}
\author[affiliation={1}]{JianHao}{Ye}
\author[affiliation={1}]{Siyu}{Hao}
\author[affiliation={1}]{Siwei}{Xia}
\author[affiliation={1},correspondingauthor]{Hongbin}{Zhou}

\address{
    $^1$ Shanghai Himalaya Technology Co., Ltd
}

\email{min1.liu@ximalaya.com, jingjing.yin@ximalaya.com, hongbin.zhou@ximalaya.com}


\usepackage{comment}
\usepackage{makecell}
\usepackage{multirow}
\usepackage{CJKutf8}
\usepackage{tcolorbox}
\usepackage{newunicodechar}
\newunicodechar{，}{,}
\usepackage{hyperref}
\usepackage[font=footnotesize,skip=2pt]{caption}
\begin{document}

\maketitle

\begin{abstract}
    Existing text-to-speech systems predominantly focus on single-sentence synthesis and lack adequate contextual modeling and fine-grained control for coherent multicast audiobooks. To address this, we propose a context-aware, emotion-controllable speech synthesis framework with three innovations: a context mechanism for consistency, a disentanglement paradigm to decouple style from prompts, and self-distillation to boost expressiveness. Experiments show consistent gains: chapter-level generation achieves 4.25 M-MOS(14\% relative gain over the strongest baseline), dialog reaches 4.11 S-MOS, and emotion control improves by 31 percentage points in high-intensity discrimination. Ablation studies validate our methods. Demo: https://semisemi-ux.github.io/.
\end{abstract}

\section{Introduction}

\label{sec:intro}

Long-form audiobooks, a widely adopted content format, integrate information delivery and auditory engagement. However, traditional production methods, whether fully manual or human-involved, face high costs and long production cycles, especially for multicast albums with multiple characters.



In response, researchers have recently developed automated solutions for high-quality audiobook generation\cite{park2025multiactor,guo2025audiostory,rong2025dopamine,xu2025mm,selvamani2025multiagent}. AudioStory\cite{guo2025audiostory} employs an LLM to process instruction inputs, decomposes long audio into structured subtasks, and generates short clips sequentially. MultiActor-Audiobook\cite{park2025multiactor} utilizes a Transformer-based multimodal model to capture character traits, uses LLMs for emotional guidance, and synthesizes speech at the sentence level. Dopamine Audiobook\cite{rong2025dopamine} and MM-StoryAgent\cite{xu2025mm} adopt multi-agent pipelines for story-based generation, but integrate existing TTS systems such as cosyvoice2\cite{du2024cosyvoice} rather than proposing new synthesis methods. Similarly, Shaja et al.\cite{selvamani2025multiagent} proposed a system that enhances immersion via spatial audio, which also relies on established TTS backbones. These approaches achieve better automation but are limited by insufficient explicit inter-sentence modeling, consequently yielding inconsistent contextual outputs.

Several industry approaches have incorporated context modeling for long-form speech. MoonCast\cite{ju2025mooncast} targets podcast generation with long-context modeling and colloquial scripts; MOSS-TTSD\cite{moss2025ttsd} achieves state-of-the-art long-segment quality through efficient codecs and data pipelines; CoVoMix\cite{zhang2024covomix} and koel-TTS\cite{hussain2025koel} focus on conversational expressiveness. Despite these advances, such systems remain largely tailored to podcasts, modeling long context with limited controllability---a critical requirement for audiobooks, which demand precise narrative flow and expressive multi-character portrayal.

Some prior efforts partially address context or controllability but exhibit inherent shortcomings. The prosody analysis in\cite{pethe2023prosody} offers data without synthesis capabilities; TACA-TTS\cite{guo2024text} improves long-sequence continuity but lacks emotional control; the memory module in\cite{li2025longcontext} over-relies on extended context, which we find can introduce redundancy and persona inconsistency. cosyvoice2\cite{du2024cosyvoice} uses instructed data to improve controllability, yet it is still short of scene adaptation for audiobooks.

To overcome these limitations, we propose a novel speech synthesis framework tailored for long-form multicast audiobooks, named Audiobook-CC. Our main contributions include the following.
\begin{itemize}
    \setlength{\itemsep}{0pt}
    \setlength{\parsep}{0pt}
    \setlength{\topsep}{0pt}
    \setlength{\partopsep}{0pt}
    \item We introduce a context modeling mechanism for long-form audiobook generation that significantly improves semantic consistency across context.
    \item A disentanglement training paradigm is designed to decouple the style of generated speech from speech prompt, which facilitates the tone of generated speech following the semantic information of the given text more closely.    
    \item A self-distillation method is proposed to greatly enhance the emotional expressiveness and controllability of the instruction of the generated speech.
\end{itemize}

\section{METHODOLOGY}

\label{sec:format}

\begin{figure*}[t] 
  \centering
  \includegraphics[width=0.96\textwidth]{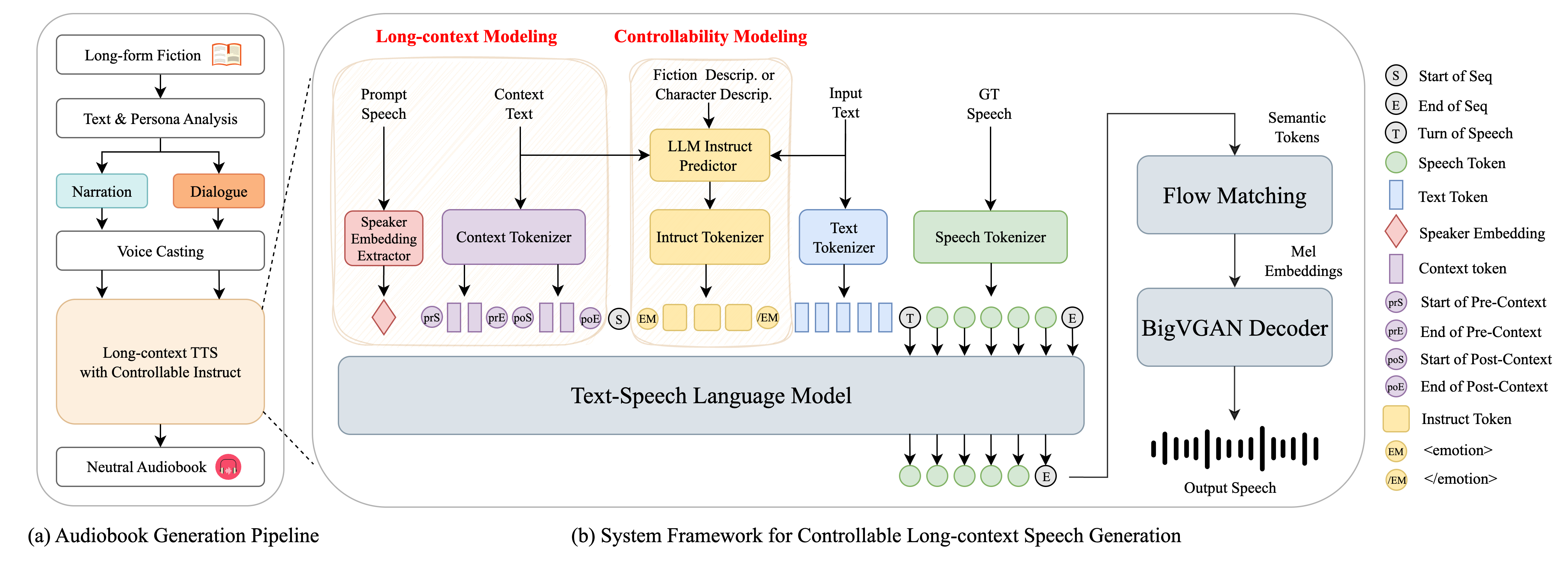}
  \caption{An overview of Audiobook-CC. (a) The audiobook generation pipeline: long-form fiction is segmented into chapters, followed by textual and character persona analysis. Narrations and dialogues are extracted, assigned to voices through casting, and synthesized using the proposed architecture, resulting in a neutral-style multicast audiobook. (b) Detailed architecture of the proposed controllable long-context speech synthesis model.}
  \label{overview}
\end{figure*}

We address four key challenges in long-form multicast audiobook synthesis: contextual scenario adaptability, semantic-prosodic consistency, character persona stability, and controllability. To tackle these challenges, we propose the Audiobook-CC system, an integrated audiobook generation pipeline illustrated in Fig. \ref{overview}(a).

The main architecture of our proposed speech generation model shown in Fig.\ref{overview}(b) leverages cosyvoice2\cite{du2024cosyvoice}. We preserve its text\&speech tokenizer and flow-matching modules, but notably substitute the HiFi-GAN vocoder with BigVGAN to enhance audio fidelity and robustness. Our work focuses mainly on exploring how to better utilize contextual information and enhance controllability in the training process of large language models. 
For auto-regressive LLMs, the organization of sequential structures is of critical importance. In our system, it is constructed as follows:

\begin{equation}
    \centering
    \footnotesize
    \begin{split}
        \mathsf{[V, seqC, S, seqE, \{W^i\}, T, \{U^j\}, E]}
    \end{split}
\end{equation}



The symbol $\mathsf{V}$ in equation$\mathsf{(1)}$ denotes a speaker embedding—not speech tokens or waveform features—and is concatenated directly to the LLM input. Crucially, this prompt embedding $\mathsf{V}$ is extracted from an utterance spoken by the same target speaker but with semantically unrelated content. This design ensures consistent vocal timbre and character persona while deliberately eliminating prosodic carryover from the prompt. Consequently, the synthesized prosody of the target segment depends solely on its own textual semantics and contextual cues, not on the rhythm, intonation, or emotional tone of the prompt audio. 

And the symbols $\mathsf{S}$ and $\mathsf{E}$ denote the start and end of a sequence, while $\mathsf{T}$ marks the switch between text tokens and speech tokens. Text tokens $\mathsf{W^i}$ are obtained by processing the input text with a text tokenizer. Similarly, the speech tokens $\mathsf{U^j}$ are derived using the same speech tokenizer in cosyvoice2\cite{du2024cosyvoice}.
An embedding of the speaker $\mathsf{V}$, derived from Cam++~\cite{wang2023cam}, is added to the input. Contextual information is structured into pre-context $\mathsf{C_{pre}}$ and post-context $\mathsf{C_{post}}$, delimited by $\mathsf{preS}$, $\mathsf{preE}$ and $\mathsf{poS}$, $\mathsf{poE}$, respectively. At inference time, post-context is obtained from the following sentence in the known audiobook script; at chapter boundaries, an empty post-context marker is used.
The full contextual sequence is formed by concatenating pre-context and post-context:

\begin{equation}
\footnotesize
\begin{split}
    \mathsf{seqC = (\underbrace{preS}_{pre-start}, \{C_{pre}^m\}, \underbrace{preE}_{pre-end}, \underbrace{poS}_{post-start}, \{C_{post}^n\}, \underbrace{poE}_{post-end})}
\end{split}
\end{equation}
Control inputs (e.g., emotion or volume) are denoted $\mathsf{E^l}$, with $\mathsf{EM}$ and $\mathsf{\backslash EM}$ marking the boundaries. The controllable sequence is constructed as follows.

\vspace{-4mm}
\begin{equation}
    \centering
    \footnotesize
    \begin{split}
        \mathsf{seqE = (EM,\{E^l\},\backslash EM)}
    \end{split}
\end{equation}

We train an autoregressive speech language model(AR) conditioned on the speaker embedding $\mathsf{V}$, $\mathsf{seqC}$, controllable $\mathsf{seqE}$, text tokens $\mathsf{W^i}$, and GT tokens $\mathsf{U^j}$, to predict the shifted speech token sequence $\mathsf{\tilde{U_j}}$, formulated as:


\vspace{-2mm}
\begin{equation}
\footnotesize
\mathsf{\prod_{j=1}^{N} p(\tilde{U}_{j} \mid V,seqC,seqE,W,\tilde{U}_{<j}; \theta_{\mathrm{AR}})}
\end{equation}
We describe in more detail the training and inference strategies of our system in the following.

\subsection{Contextual Consistency via Disentangled Training}
Although pre-context $\mathsf{C_{pre}}$ and post-context $\mathsf{C_{post}}$ intuitively enhance contextual consistency, training solely on textual context without explicit speaker modeling still requires prompt speech for voice specification during inference. Consequently, synthesized speech prosody remains constrained by the prompt speech, limiting semantic-prosodic alignment and narrative coherence in long-form synthesis.

\noindent\hspace{0.5cm}\textbf{semantic consistency:} To improve the alignment between synthesized audio and text semantics while mitigating excessive prosodic interference from prompt audio, we adopt a decoupled training strategy. Rather than the traditional coupled ``prompt-target'' training model, we employ independent modeling that focuses on the adaptive relationship between target audio prosody and text semantics. During training, for example, we use the audio1 of Speaker A (with unrelated content) as the source of V to synthesize the audio2 of Speaker A, guiding the model to preserve shared speaker traits—such as timbre and persona—while allowing the prosody and expressiveness of audio2 to emerge purely from its corresponding text and context. A key practical challenge in realizing this approach lies in acquiring large-scale, high-expressiveness datasets where individual speakers are represented with sufficient data (each speaker covers no less than 5 minutes of data) spanning diverse narrative styles, emotions, and speaking rates. This approach reduces inter-audio prosodic interference at the data level and enhances consistency between synthesized audio and current text semantics.

\noindent\hspace{0.5cm}\textbf{persona consistency:} To avoid acoustic timbre shifts caused by insufficient prompt matching while decoupling, we propose a multi-constraint prompt selection mechanism with three refinements: 1) prompts are selected within the same chapter with voiceprint similarity constraints~\cite{wang2023cam} to reduce cross-chapter variance; 2) an optimal similarity threshold is experimentally determined to balance voice stability and prosodic diversity; 3) high-quality labeled multi-emotional speech data is incorporated to alleviate sample sparsity and ensure consistent persona expression across emotions.

\subsection{Controllability Enhancement}

This section introduces our controllability enhancement strategy, focusing on training and inference mechanisms.

\noindent \hspace{0.5cm}\textbf{controllability:}
During training, instructions derived from fiction or persona are decomposed into discrete attribute labels. Based on datasets with high prosodic consistency, we use LLM to parse instructions or contextual text, extracting nine emotions (admire/alert/anger/disgust/fear/joy/neutral/sadness/surprise) along with their intensity levels, as well as categorical values for volume (high/medium/low) and speaking rate (fast/medium/slow). The LLM output is then normalized by manually defined rules into fixed tokens, where emotion intensity is quantized into four standardized levels: extremely, considerable, some, and slight. This enables the generation of rich and compositional control signals such as ``very joyful with slight surprise''.  For example, ``shouting angrily'' decomposes into ``very angry, low volume, slow speed'' for an ill elderly person and ``very angry, high volume, fast speed'' for a healthy person, facilitating explicit attribute--acoustic feature mapping learning. During inference, a third module converts user instructions into attribute combinations. This discretization mitigates linguistic ambiguity and improves the precision of control.

\noindent \hspace{0.5cm}\textbf{emotional intensity:} To alleviate the scarcity of high-intensity emotional samples,  we first trained an initial model using data collected internally from character-voice. Subsequently, the synthetic data generated by this initial model—with qualified samples(the phoneme error rate(PER)\cite{smith2019per} $\mathsf{<}$ 2\%, speaker similarity (SS)\cite{wang2023cam} $\mathsf{>}$ 0.7 and pitch fluctuation consistent with emotional characteristics) and highly emotional data prioritized as targeted data—is used to iteratively refine the stability of the model. This process employs a self-distillation strategy consisting of three key steps:
First, samples with varying intensity levels are synthesized using a pre-trained emotional TTS model;
Second, the generated samples are filtered through the phoneme error rate (PER)\cite{smith2019per}, speaker similarity (SS)\cite{wang2023cam} and pitch to guarantee quality;
Third, the intensity distribution is balanced through targeted data augmentation.





\section{Experimental Setup}

\label{sec:pagestyle}

\subsection{Experimental Setup}
\noindent \hspace{0.5cm}\textbf{Datasets and Training:} There were three stages in our training. First, 1,000,000 hours of data was used—including audiobooks, podcasts, TV dramas, etc.—to fine-tune the model initialized from cosyvoice2\cite{du2024cosyvoice}. Second, a 150,000-hour context-aware audiobook dataset and a 100-hour dataset of high-quality recordings labeled with instructions were used to further fine-tune the model. In particular, each sentence in the context-aware data set was annotated with timestamps, speaker IDs, and positional information within the corresponding chapter. Third, to enhance the expressiveness and controllability of instruction-following, a 100-hour data augmentation dataset based on the model constructed on the model after the first two training stages.

During training, the model was optimized using the AdamW optimizer\cite{loshchilov2017decoupled}, with learning rates of 1e-4, 1e-5, and 1e-6 for the first, second and third stages, respectively. The model was trained on 64 NVIDIA A800 GPUs with a batch size of 384 for 720k, 300k, and 10k steps in the three stages, respectively.

To evaluate performance in audiobook scenarios, we constructed three test sets: a narrative test set (Test-NAR) consisting of 100 paragraphs (each containing more than 240 sentences), a dialog test set (Test-DIA) comprising 570 dialog sentences, and a chapter-level test set (Test-CHAP) built from 15 chapters (each ranging from 2,000 to 4,000 sentences) that include both narration and dialog.

\noindent \hspace{0.5cm}\textbf{Model Evaluation: } For a comprehensive evaluation of our system, we employed two subjective evaluation methods\cite{lei2023contextAware}: the single-sentence mean opinion score (S-MOS) for dialog samples and the multi-sentence mean opinion score (M-MOS) for narration and long-chapter samples. We randomly selected 20 paragraphs from \textit{Test-NAR}, 60 sentences from \textit{Test-DIA}, and 10 chapters from \textit{Test-CHAP} for subjective rating. Fifty native Chinese speakers were recruited to evaluate both S-MOS and M-MOS.

Furthermore, we conducted ABX listening tests to compare our system against two baselines: single-sentence ABX (S-ABX) for dialog samples and multi-sentence ABX (M-ABX) for narration and chapter-level samples. The baselines---cosyvoice2\cite{du2024cosyvoice} and MOSS-TTSD\cite{moss2025ttsd}---were selected to cover distinct synthesis regimes: cosyvoice2\cite{du2024cosyvoice} demonstrates strong stability and expressiveness in short-form synthesis, whereas MOSS-TTSD\cite{moss2025ttsd} has shown effectiveness in long-form speech generation. Each baseline is evaluated using its native recommended configuration.

In addition to subjective evaluations, we employed objective metrics—phoneme error rate (PER)\cite{smith2019per} and speaker similarity (SS)\cite{wang2023cam}—to evaluate phonetic precision and speaker consistency of our model. Our proposed model was evaluated and compared under the following three inference configurations:
\textit{Infer-ctx}: inference using only the additional context sequence;
\textit{Infer-inst}: inference using only the additional instruction sequence;
\textit{Infer-ctx\&inst}: inference incorporating both context and instruction sequences.


\subsection{Main Results}
\begin{figure}[h]
\centering
\footnotesize
    \includegraphics[height=3.35cm]{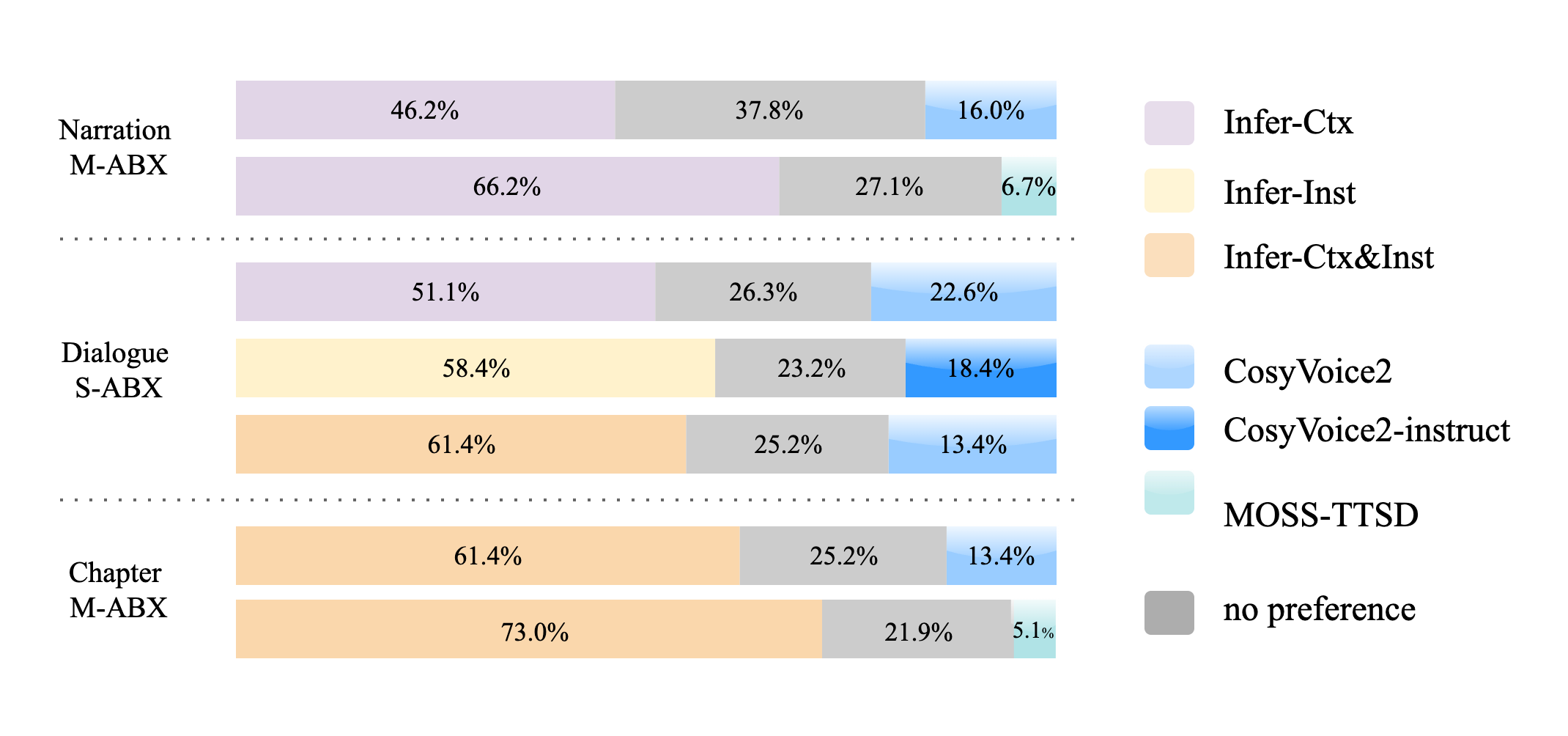}
	\caption{\label{abx_test} {Comparison of preference among different models}}
\end{figure}

\vspace{-2mm}
The results of Table \ref{t_mos_overall} and Fig.\ref{abx_test} show that the proposed system largely exceeds both baseline models for all test setups. By comparing the results between the narration and dialog tests, the results of the dialog tests are more advantageous than those of the narration tests for both MOS and ABX tests.
\begin{table}[ht]
\centering
\footnotesize
\caption{The S-MOS and M-MOS of different context models}
\setlength{\tabcolsep}{4pt}
\renewcommand{\arraystretch}{0.8} 
\begin{tabular}{lccc}
\hline
\textit{Model} & \textit{\makecell[c]{Narration\\M-MOS}} & \textit{\makecell[c]{Dialogue\\S-MOS}} & \textit{\makecell[c]{Chapter\\M-MOS}} \\
\hline
\addlinespace[1pt] 
\textit{Baseline} \\
\quad cosyvoice2\cite{du2024cosyvoice} & 3.91±0.07 & 3.84±0.09 & 3.88±0.07 \\
\quad MOSS-TTSD\cite{moss2025ttsd} & 3.78±0.11 & 3.67±0.07 & 3.72±0.09 \\
\hline
\addlinespace[1pt] 
\textit{Proposed} \\
\quad Infer-ctx & \textbf{4.06±0.08} & 3.93±0.06 & 4.13±0.09 \\
\quad Infer-inst & / & 3.96±0.08 & / \\
\quad Infer-ctx\&inst & / & \textbf{4.11±0.06} & \textbf{4.25±0.11} \\
\hline
\label{t_mos_overall}
\end{tabular}
\end{table}
\vspace{-2mm}

It is also worth mentioning that the Infer-ctx\&inst speech generation strategy benefits \textit{Infer-ctx\&inst} by combining context-aware narration generation and instructed dialog generation. ABX preference rates comparing the inference configurations of the proposed system with baselines. Infer-ctx\&Inst achieves the strongest performance (\textit{73.0\%} chapter-level, \textit{61.4\%} dialog), demonstrating synergistic benefits of combining long-context modeling with instruction-based control. Context-aware inference alone (Infer-ctx) substantially outperforms baselines in narration (\textit{66.2\%}) and chapter-level (\textit{61.4\%}) scenarios, highlighting the importance of contextual consistency for long-form generation.

\subsection{Ablation Analysis}

To validate key components of our framework, four ablation experiments were conducted. Experiments 1–3 focus on context awareness, using a 150 k-hour audiobook chapter data set for training, Test-DIA for evaluation, and phoneme error rate (PER)\cite{smith2019per} and speaker similarity (SS)\cite{wang2023cam} and S-MOS as unified metrics. Experiment 4 focuses on instruction, adopting dedicated training data—including 500 hours of high-quality emotional data and 5,300 hours of augmented data—and tailored assessment methods.

\noindent\hspace{0.5cm}\textbf{Decoupled vs Non-Decoupled Models:} We compared two strategies for the context-aware module: the non-decoupled model, where audio prompts and targets are identical, and the decoupled model, in which distinct prompts and targets are selected via a specific strategy. The results in Table \ref{context strategy} show that the non-decoupled model achieves excessively high speaker similarity (SS)\cite{wang2023cam}, indicating over-similar timbre, prosody, and emotion—and thus impractical for an audiobook due to insufficient character diversity.

\noindent\hspace{0.5cm}In "Decoupled-0.68/0.8", both 0.68 and 0.8 represent the speaker embedding similarity between the prompt audio and the target audio during the decoupled training paradigm, which is used to balance speaker consistency and emotional diversity.

\begin{table}[ht]
\centering
\footnotesize
\caption{Performance Comparison of Context Strategy}
\setlength{\tabcolsep}{4pt}
\renewcommand{\arraystretch}{0.8} 
\label{context strategy}
\begin{tabular}{lccc}
\hline
\textit{Model} & \textit{SS} & \textit{PER} & \textit{S-MOS}  \\
\hline
Non-Decoupled & \textbf{0.87} & 1.33 &  3.45±0.09 \\    
Decoupled-0.68 & 0.69 & 1.19 & 3.86±0.06 \\
Decoupled-0.8 & 0.79 & 1.84 & 3.82±0.07 \\
Decoupled-0.8 + context & 0.80  & 1.67 & \textbf{3.93±0.06} \\
\hline
\end{tabular}
\end{table}
\noindent\hspace{0.5cm}\textbf{Effect of Decoupling Threshold:} Audio clips within chapters were clustered using different thresholds to generate distinct speaker IDs. A lower threshold reduced speaker similarity (SS)\cite{wang2023cam}, with occasional timbre discontinuities, but yielded slightly higher S-MOS than higher thresholds. Conversely, an excessively high threshold is hypothesized to increase speaker similarity (SS)\cite{wang2023cam}, approaching non-decoupled model performance, while risking lower S-MOS, based on tested threshold trends.

\noindent\hspace{0.5cm}\textbf{Effect of Contextual Text Input:}
We evaluated the influence of contextual text on the context-aware module by feeding the Text-Speech Language Model with two input types: target text alone, or target text plus its preceding and subsequent sentence. The context-augmented model achieved a higher S-MOS, with listening tests confirming improved coherence. Next, we present an example.

\begin{tcolorbox}[
    colback=gray!8,
    colframe=gray!20,
    boxrule=0.1pt,
    arc=1pt,
    top=1pt,
    bottom=1pt,
    left=2pt,
    right=2pt,
    before skip=1pt,
    after skip=1pt
]

\begin{CJK*}{UTF8}{gbsn}
\scriptsize``叫师父多给你喂几碗符水，看你还胡说八道。我猜是师兄功力大进''

While this sentence shows no explicit emotional indicators, the model naturally incorporates laughter into the synthesis based on the prior context ``一旁的夏姝噗嗤笑个不停，''
\end{CJK*}
\end{tcolorbox}

\noindent \hspace{0.5cm}\textbf{Fine-Grained Emotional Control: } We used the Chinese emotional speech test set from CV3-Eval\cite{du2025cosyvoice} to evaluate controllability, modifying the original instructions into three types of emotional states: ``high-intensity single emotion'', ``low-intensity single emotion'', and ``mixed emotion''. Three metrics are used for emotional evaluation: emotion classification score F1 \cite{ma2023emotion2vec} for single emotions; S-MOS\cite{lei2023contextAware} and speaker similarity (SS)\cite{wang2023cam} for mixed emotions. The results in Table \ref{Single Emotion} show that our model exhibits stronger H-L (discriminability between ``high-intensity'' and ``low-intensity'') emotion control in the ``Text-Unrelated'' test set and outperform baseline models\cite{du2024cosyvoice} in high-intensity emotion control. Table \ref{Mixed Emotion} shows that our model also outperforms the baseline model\cite{du2024cosyvoice} in terms of mixed emotion performance.

In Table 5, we compare the impact of self-distilled data on model stability. The results show that incorporating self-distilled data significantly improves inference stability, primarily reflected in a reduction in phoneme error rate (PER). The decrease in F1 in the 700-hour subset is mainly caused by distribution differences between synthetic and human-recorded data; with more balanced self-distilled data, the emotion F1 recovers while PER remains lower.

\begin{table}[!ht]
\centering
\scriptsize
\caption{Comparison of F1 scores For Single Emotion}
\label{Single Emotion}
\setlength{\tabcolsep}{4pt}
\renewcommand{\arraystretch}{0.74} 
\begin{tabular}{@{}l|*{3}{c}|*{3}{c}@{}}
\toprule
\multirow{2}{*}{\textit{Model}} & \multicolumn{3}{c|}{\textit{Text-Related}} & \multicolumn{3}{c}{\textit{Text-Unrelated}} \\
\cmidrule(lr){2-4} \cmidrule(lr){5-7}
 & \textit{angry} & \textit{happy} & \textit{sad} & \textit{angry} & \textit{happy} & \textit{sad} \\
\midrule
cosyvoice2-H & 0.79 & 0.96 & 0.68 & 0.07 & 0.62 & 0.53 \\
cosyvoice2-L & 0.77 & 0.92 & 0.68 & 0.00 & 0.68 & 0.38 \\
Infer-inst-H & 0.72 & 0.92 & 0.98 & 0.31 & 0.54 & 0.65 \\
Infer-inst-L & 0.62 & 0.91 & 0.94 & 0.00 & 0.39 & 0.32 \\
\midrule
$\Delta_{\text{cosyvoice2}}$(H-L) & 0.02 & 0.04 & 0.00 & 0.07 & -0.06 & 0.15 \\
$\Delta_{\text{Infer-inst}}$(H-L) & \textbf{0.10} & 0.01 & \textbf{0.04} & \textbf{0.31} & \textbf{0.15} & \textbf{0.33} \\
\bottomrule
\end{tabular}
\end{table}
\vspace{1mm}

\begin{table}[!ht]
\centering
\scriptsize
\caption{Performance Comparison For Mixed Emotion}
\label{Mixed Emotion}
\setlength{\tabcolsep}{4pt}
\renewcommand{\arraystretch}{0.74} 
\begin{tabular}{@{}l|*{2}{c}|*{2}{c}@{}}
\toprule
\multirow{2}{*}{\textit{Model}} & \multicolumn{2}{c|}{\textit{Text-Related}} & \multicolumn{2}{c}{\textit{Text-Unrelated}} \\
\cmidrule(lr){2-3} \cmidrule(lr){4-5}
 & \textit{SS} & \textit{S-MOS} & \textit{SS} & \textit{S-MOS} \\
\midrule
cosyvoice2-instruction & 0.74 & 3.67±0.06 & 0.75 & 3.35±0.07 \\
Infer-inst & \textbf{0.77} & \textbf{4.08±0.07} & \textbf{0.78} & \textbf{3.87±0.09} \\
\bottomrule
\end{tabular}
\end{table}
\vspace{1mm}

\begin{table}[!ht]
\centering
\scriptsize
\caption{Performance Comparison of Different Self-Distillation Strategies (Text-Related Metrics)}
\label{tab:mixed_emotion_text_related}
\setlength{\tabcolsep}{4pt}
\renewcommand{\arraystretch}{0.74} 
\begin{tabular}{@{}l c c c c c@{}}
\toprule
\textit{Model} & \textit{csft\_data} & \textit{PER(\%)} & \textit{Angry} & \textit{Happy} & \textit{Sad} \\
\midrule
benchmark       & rec\_100h                & 4.54 & 0.73 & 0.89 & 0.96 \\
model-700h      & rec\_100h+700h   & 2.37 & 0.65 & 0.72 & 0.81 \\
model-1400h     & rec\_100h+1400h  & 1.26 & 0.70 & 0.90 & 0.94 \\
model-3000h     & rec\_100h+3000h  & 1.23 & 0.72 & 0.92 & 0.98 \\
\bottomrule
\end{tabular}
\end{table}

\section{Conclusions}
\label{sec:majhead}
The paper proposes a controllable, context-aware TTS framework for multicast audiobooks, with three innovations: 1) a context mechanism for contextual consistency, 2) a disentanglement paradigm to decouple style control from speech prompts for semantic consistency, and 3) self-distillation to boost emotional expressiveness and controllability.
Experiments show that the framework outperforms baselines in narration, dialog, and chapter generation, with ablation studies validating its key components. In the future, we can expand the chapter data or select specific data to mitigate data sparsity, explore reinforcement learning for performance improvement, and further investigate ``thinking''-style architectures for speech models, where contextual understanding and model control are directly trained within the large speech generative model itself, rather than being delegated to external or modular components.

\section{Use of Generative AI Disclosure}
Generative AI tools are employed in this work solely for language editing, polishing, and manuscript formatting. They are not used to generate any core content, research ideas, experimental designs, results, or major textual components of the paper. All scientific contributions, including model design, experiments, analysis, and conclusions, are fully completed by the authors.

\bibliographystyle{IEEEtran}
\bibliography{interspeech_2026}
\end{document}